\documentclass{article}
\usepackage{longtable}
\usepackage[utf8]{inputenc}
\usepackage[T1]{fontenc}
\usepackage{graphicx}
\usepackage{hyperref}
\usepackage{xcolor}
\usepackage{listings}
\usepackage{caption}
\usepackage{array}
\usepackage{fullpage}

\title{Browser Security Posture Analysis: A Client-Side Security Assessment Framework}
\author{
  Avihay Cohen \\[1ex]
  \small Seraphic Algorithms
}
\date{May 11, 2025}

\begin{document}
\maketitle

\begin{abstract}
Modern web browsers have effectively become the new operating system for business applications, yet their security posture is often under-scrutinized. This paper presents a novel, comprehensive \textbf{Browser Security Posture Analysis Framework}\cite{browser_security_framework}, a browser-based client-side security assessment toolkit that runs entirely in JavaScript and WebAssembly within the browser. It performs a battery of over \textbf{120 in-browser security tests} in situ, providing fine-grained diagnostics of security policies and features that network-level or os-level tools cannot observe. This yields insights into how well a browser enforces critical client-side security invariants. We detail the motivation for such a framework, describe its architecture and implementation, and dive into the technical design of numerous test modules (covering the same-origin policy, cross-origin resource sharing, content security policy, sandboxing, XSS protection, extension interference via WeakRefs, permissions audits, garbage collection behavior, cryptographic APIs, SSL certificate validation, advanced web platform security features like \texttt{SharedArrayBuffer}, Content filtering controls ,and internal network accessibility). We then present an experimental evaluation across different browsers and enterprise scenarios, highlighting gaps in legacy browsers and common misconfigurations. Finally, we discuss the security and privacy implications of our findings, compare with related work in browser security and enterprise endpoint solutions, and outline future enhancements such as real-time posture monitoring and SIEM integration.
\end{abstract}

\section{Introduction}
Web browsers are no longer simple rendering tools for static pages; they have evolved into de facto operating systems for modern applications. In today’s enterprise environments, employees often spend the majority of their workday inside a browser, using web-based SaaS applications for critical business functions. One study found that nearly 90\% of an organization’s work is conducted in the web browser. With this shift, browsers have become a primary interface to sensitive corporate data and services. Consequently, the browser’s security posture – its configuration, policies, and runtime behavior – has a direct impact on enterprise security.

However, traditional network-centric/os-centric security architectures (firewalls, secure web gateways, proxies,EDR, etc.) were not designed with this browser-centric reality in mind. These solutions focus on monitoring and filtering network traffic, os activity , but cannot fully assess or control what happens inside the browser. As traffic is increasingly end-to-end encrypted (HTTPS by default) and browsers execute rich client-side scripts, many security-critical decisions are made within the browser environment itself—beyond the visibility of network/os tools. For example, whether a file download from an in-memory blob is allowed is determined by the browser at runtime, not by the network. Enterprise security teams are finding that solely relying on perimeter defenses leaves blind spots in browser behavior.

This paper addresses the rising significance of browser security by introducing \emph{Browser Security Posture Analysis Framework}\cite{browser_security_framework}, a client-side framework for comprehensive browser security assessment. Our framework operates within the browser, using JavaScript and WebAssembly to perform a battery of security tests in situ. By living inside the “user space” of the browser, it can evaluate fine-grained behavior—such as DOM manipulation, policy enforcement, and internal API behavior—that would be invisible to external network/os monitors. In essence, Browser Security Posture\cite{browser_security_framework} treats the browser itself as the subject of scrutiny, much like a vulnerability assessment tool would treat an operating system.

We proceed as follows: Section~2 articulates the motivation and requirements for a client-side browser assessment framework in enterprise contexts. Section~3 provides an overview of the architecture and implementation of Browser Security Posture\cite{browser_security_framework}, emphasizing its self-contained, entirely in-browser design. Section~4 details each test module and the technical strategies used to evaluate various security features (ranging from classical web policies like the Same-Origin Policy to novel detection of external interference via WeakRefs). Section~5 presents an experimental evaluation with example results across multiple browsers and scenarios, illustrating the framework’s capability to highlight security posture differences. Section~6 discusses the security and privacy implications of the findings and the risks of not performing such diagnostics. Section~7 reviews related work, comparing our approach to prior browser security testing tools and enterprise solutions (such as remote browser isolation and secure enterprise browsers). Section~8 concludes with a summary and outlines future work, including potential real-time posture monitoring extensions and integration into enterprise security operations.

\section{Motivation and Requirements}
As enterprises embrace cloud-first strategies and remote work, browsers now mediate access to a vast array of corporate resources. This expanded role introduces new attack surfaces and failure modes. Users may unwittingly run browsers with insecure configurations, outdated policies, or malicious extensions, undermining corporate security from within the browser itself. Key motivations for a client-side browser security framework include:

\begin{itemize}
    \item \textbf{Invisibility of Browser Internals to Network Tools:} Many security-relevant browser decisions leave no network trace. For instance, whether an in-memory file download occurred, or whether a webpage attempted to access a local device or a sensitive API, are actions enforced entirely within the browser, beyond the view of any network monitor. These internal events could pose risks (e.g., a malicious script accessing a local resource or data) yet are undetectable by perimeter defenses. A client-side framework can directly observe and test such behavior to reveal hidden vulnerabilities.

    \item \textbf{Complementing Zero Trust and Endpoint Security:} Zero Trust models emphasize continuous verification of device and application posture. While endpoint security solutions monitor OS-level metrics, they often lack visibility into browser-specific settings and activities (like which sites have risky permissions or how the browser handles certain content). Browser Security Posture\cite{browser_security_framework} fills this gap by acting as a specialized “browser auditor” on the endpoint, complementing endpoint protection platforms (EPP/EDR) with browser-focused checks. Unlike generic endpoint agents, our framework is tailored to the browser’s execution environment and can perform targeted tests (like attempting a known bad SSL connection) safely and in a standardized manner. 
    
    \item \textbf{Detection of Misconfigurations and Policy Drift:} Browsers across a fleet may gradually deviate from baseline configurations due to user actions, unmanaged updates, or faulty GPO/MDM propagation. A client-side framework can detect such drifts in real time, ensuring policy adherence.

    \item \textbf{Protection Against Malicious Extensions and Plugins:} Extensions often operate with elevated privileges and can leak data, inject malicious scripts, or surveil user behavior. A Browser Security Posture framework can help detect extensions that inject malicious JavaScript, such as those that attempt to capture sensitive DOM elements (e.g., login forms, tokens, or clipboard contents).

    \item \textbf{Visibility into Browser-Based Data Exfiltration:} Sophisticated data exfiltration tactics, such as copy-paste monitoring, JavaScript-based uploads, or clipboard manipulation, occur entirely within the browser. These bypass both network DLP and OS-level logging unless monitored from inside the browser.

    \item \textbf{Enhanced Compliance and Auditability:} Certain regulatory standards (e.g., HIPAA, PCI DSS, ISO 27001) require monitoring of user activities and control over access to sensitive data. A browser-focused security posture tool can provide compliance evidence by logging access patterns and enforcing browser hardening guidelines.

    \item \textbf{Safe Simulation of Attacks:} A client-side framework can simulate known browser-specific exploits (e.g., outdated SSL/TLS versions, localStorage abuse, DOM-based XSS vectors) in a sandboxed way to test the browser’s resilience—something not possible with passive monitoring.

    \item \textbf{Cross-Browser and Cross-Version Insights:} Enterprise environments may host a mix of Chrome, Edge, Firefox, and even legacy browsers. A unified framework enables standardized posture assessment across heterogeneous environments, accounting for version-specific vulnerabilities and settings.

\end{itemize}

In summary, the need for a client-side browser assessment framework arises from the convergence of two trends: (1) the browser’s emergence as the primary work environment (and thus a primary target for attackers), and (2) the insufficiency of legacy, network/os-centric defenses to monitor and enforce security \emph{within} the browser. Browser Security Posture\cite{browser_security_framework} is designed to empower enterprises to “trust, but verify” the security of the browsers operating in their environment by performing checks from the inside out rather than the outside in.

\section{Framework Overview}
\subsection{Architecture}
\label{sec:architecture}
Browser Security Posture\cite{browser_security_framework} is built as a self-contained web application that can be delivered to end-user browsers (for instance, via an internal security portal or as an injected script by an IT management system). Crucially, it runs entirely within the browser – there are no external executables, plugins, or native code. The framework is implemented in JavaScript (ES2020+) and leverages standard Web APIs available in modern browsers. This design choice ensures that it is platform-agnostic (works on any OS and any modern browser) and easy to deploy (just visit a URL or open an HTML file). All test logic executes on the client side. If network access is needed (e.g., to fetch a test resource or check a certificate), those requests are initiated by the browser itself, often to preconfigured endpoints. The framework does not require any special privileges; it operates within the normal sandbox of a web page, relying only on the browser’s capabilities to introspect itself. This means it cannot do anything a regular web page couldn’t do (by design – we assess security without breaking the browser’s own security model).

\begin{figure}[ht]
\centering
\includegraphics[width=0.8\textwidth]{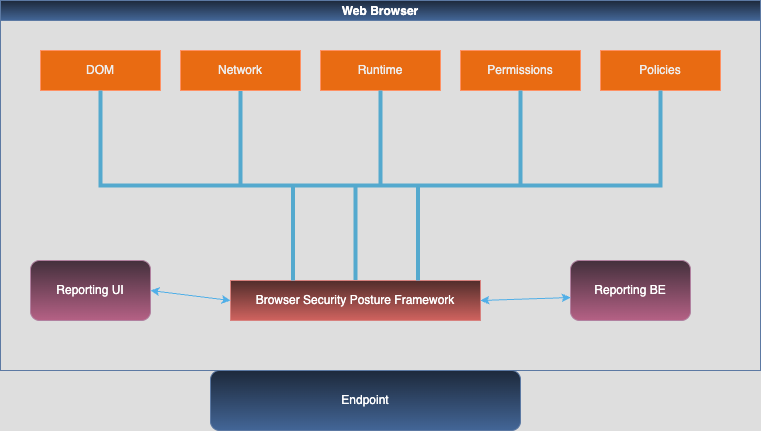}
\caption{Overview of the Browser Security Posture\cite{browser_security_framework} framework architecture and components. The in-browser agent executes a series of test modules and aggregates results for reporting.}
\label{fig:architecture}
\end{figure}

Figure~\ref{fig:architecture} illustrates the high-level architecture and flow of Browser Security Posture\cite{browser_security_framework}. The framework (running as a script inside the browser) comprises multiple test modules that examine different aspects of the browser’s security posture. These modules interact with the browser’s internal APIs, configurations, and runtime environment. Some modules may also perform controlled network requests (for example, to known test sites like \texttt{badssl.com}) to trigger browser behavior under certain conditions (such as handling of invalid SSL certificates). Throughout the testing process, the framework collects results and can present them to the user or send them to a backend for aggregation. (In an enterprise deployment, results might be reported to an IT management console or SIEM; care is taken not to exfiltrate sensitive information -- see Section~\ref{sec:privacy} on privacy considerations.) Importantly, all network interactions initiated by the framework are either loopback (targeting local network addresses for scanning) or directed to innocuous test resources. There are no external third-party script dependencies – this not only reduces supply-chain risk but also ensures consistent behavior (no CDN differences, offline capability, etc.).

\subsection{Implementation}
The core of Browser Security Posture\cite{browser_security_framework} is written in modern JavaScript, leveraging features like \texttt{async/await} for managing asynchronous test sequences and new APIs such as \texttt{WeakRef} and \texttt{FinalizationRegistry} for certain memory integrity tests (discussed later). We avoided heavy frameworks to keep the payload small. The main framework, including all tests, is approximately 200\,KB of JavaScript. It can be loaded as a single HTML file containing inline scripts for each module, or as a set of bundled modules. No external network access is required except for specific tests that intentionally fetch resources (all such network tests use well-defined URLs and can be pointed to either public test servers or an enterprise’s own test endpoints). For example, we use known hostnames like \texttt{expired.badssl.com} to test certificate validation; these hostnames can be changed to internal sites if an organization prefers an offline test mode.

The framework includes a simple UI that reports the results of each test (e.g., “CSP Enforcement: PASSED”, “WeakRef DOM Leakage: DETECTED/NOT DETECTED”). In an enterprise scenario, this UI could be suppressed and results automatically sent to a central collector. We designed the tests to run efficiently without significantly disrupting the user’s browsing session – tests run quickly (most complete in milliseconds; some longer ones like randomness analysis or network scans may take a few seconds but can be throttled or run in the background). Each module is careful to clean up after itself (removing any test DOM elements, closing connections, etc.) to avoid side effects in the browser.

\begin{figure}
    \centering
    \includegraphics[width=0.5\linewidth]{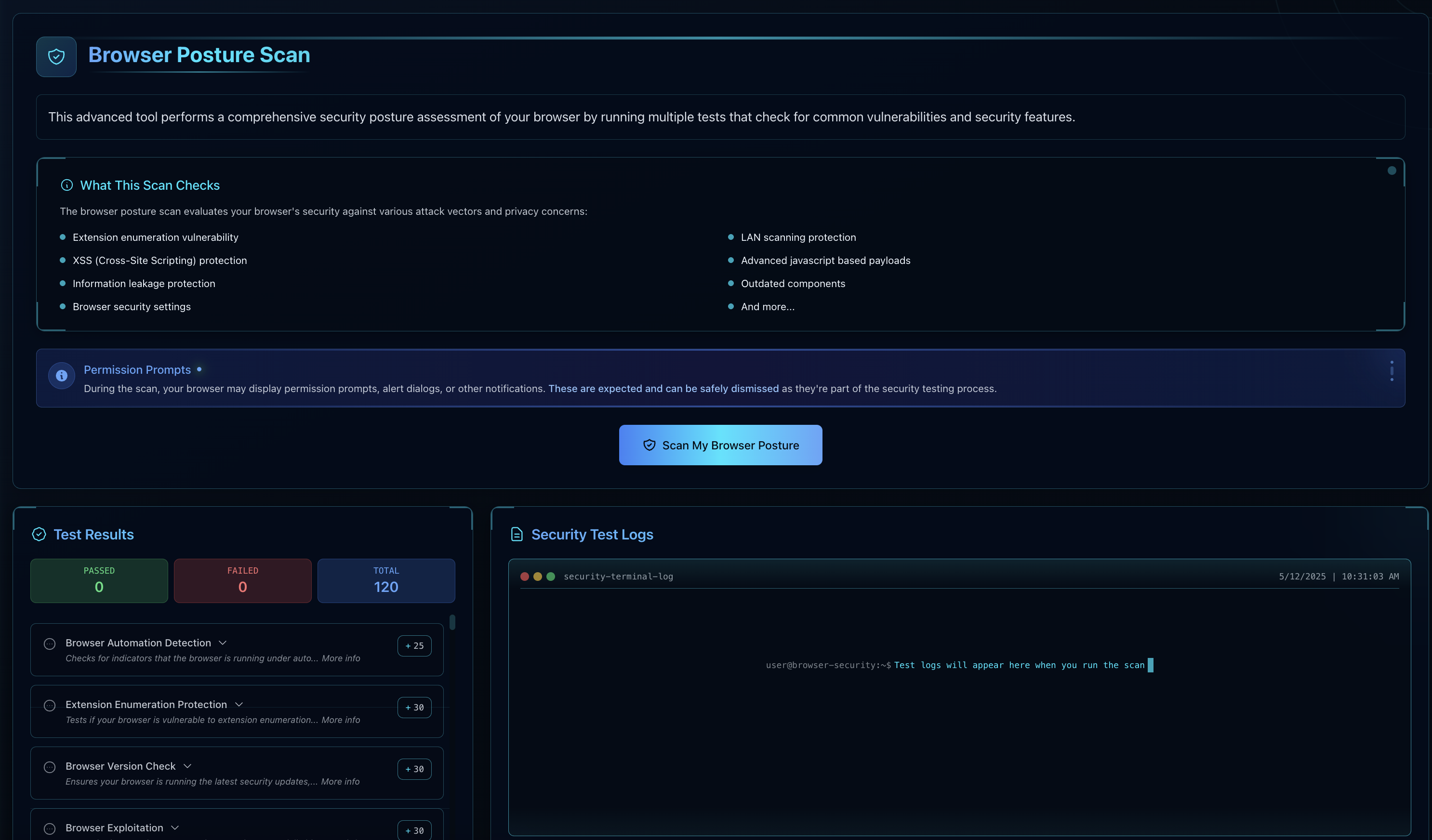}
    \caption{Browser Security Posture UI}
    \label{fig:enter-label}
\end{figure}

\subsection{Security Considerations}
Because the framework runs inside the browser, we carefully considered the possibility that malware or malicious extensions could detect or interfere with it. If a user’s browser is already compromised (e.g., by a malicious extension or a tampered binary), an advanced adversary could theoretically alter the testing script’s execution or results. Our framework includes self-integrity checks (for instance, verifying the script text against a known hash, if provided, though this is optional) and uses multiple approaches to detect interference. For example, one test module uses a \textbf{WeakRef-based DOM probe} (discussed in Section~\ref{sec:weakref}) which can reveal if an extension is trying to stealthily intercept or hold references to our test objects. Nonetheless, the framework is primarily a diagnostic tool and not an active defense mechanism; it assumes a mostly honest execution environment and is aimed at identifying misconfigurations or unnoticed vulnerabilities rather than fighting active malware head-on. When integrated into an enterprise IT workflow, running these tests regularly (e.g., at login) can alert administrators to issues needing remediation (such as “User X has an extension that retains a DOM reference – possibly unsanctioned” or “Browser Y did not enforce CSP properly”).

\section{Test Modules and Technical Details}
The framework’s test suite is organized into modules, each focusing on a different aspect of browser security. The modules can be run in sequence or in parallel, and each produces a result (pass/fail or data for analysis). We describe each major module in depth below, explaining the technical approach and any novel techniques employed.

\subsection{Same-Origin Policy (SOP) and CORS Enforcement}
\paragraph{Same-Origin Policy (SOP):} SOP is the foundational security policy of web browsers, which isolates content from different origins. It ensures that scripts from one origin (defined by scheme, host, and port) cannot freely access resources from another origin. This prevents, for example, a malicious site from reading data from a user’s webmail account open in another tab. To verify SOP enforcement, our framework conducts a series of checks: we attempt forbidden actions such as reading cross-origin DOM content, accessing cross-origin cookies, or making fetch/XHR requests to a cross-origin resource without permission. For instance, we embed an \texttt{<iframe>} from a different domain and then try to read its \texttt{document.title} or modify its content from the parent page’s script. A compliant browser should block such attempts and throw a security exception. We catch any exceptions and mark the test as passed if the access is correctly prevented (and failed if, against expectations, we succeed – which would indicate a serious SOP violation). Additionally, we test nuanced SOP corner cases (e.g., interactions with the \texttt{document.domain} relaxation setting and location hash navigation between frames) to ensure no misconfiguration or policy bypass is present. As expected, all modern browsers strictly enforce SOP. A legacy or atypical configuration where SOP is weakened would be flagged by these tests.

\paragraph{Cross-Origin Resource Sharing (CORS):} CORS is a controlled relaxation of SOP, allowing a web page to request resources from another domain if the server explicitly consents via specific HTTP headers. We test the browser’s CORS implementation by performing fetches or XHRs to a test server under various conditions. For example, we attempt to fetch a JSON resource from \texttt{example.com} while our test page is on \texttt{test.com}. The test server can be configured to respond with no CORS headers, or with \texttt{Access-Control-Allow-Origin: *} or a specific origin. Our framework verifies that the browser obeys the rules: a request without the proper response headers should be blocked by the browser (the fetch promise is rejected with a TypeError due to CORS). We try both “simple” requests and those requiring a preflight (e.g., using a custom header or the \texttt{PUT} method), ensuring the browser properly sends an OPTIONS preflight and handles the response. The expected outcome is that Chrome, Firefox, Edge, and Safari all pass, enforcing CORS correctly by blocking unauthorized cross-origin responses and allowing those with correct headers. The test logs detail each sub-case (e.g., “simple request allowed by server – should succeed; request not allowed by server – should fail”), providing fine-grained assurance that the policy holds. If a corporate environment had disabled standard web security checks, these tests would immediately reveal the misconfiguration.

\subsection{Content Security Policy and XSS Protections}
\paragraph{Content Security Policy (CSP):} CSP is a modern security standard that allows web administrators to declare allowed sources of content, serving as a mitigation against cross-site scripting (XSS) and other injection attacks. For example, a CSP might forbid inline scripts or loading scripts from unauthorized domains. To evaluate CSP enforcement, our framework needs a context where CSP is active. We employ two strategies: (1) if the framework’s host page is served with a CSP header, we can directly test policy enforcement in that context; (2) if not, we dynamically create a test context with a CSP (for instance, by injecting a \texttt{<meta>} CSP tag or opening a test page with a CSP header) and then attempt to violate the policy. In practice, we try to inject or execute scripts in ways that the CSP should forbid (such as inserting a \texttt{<script>} element with an inline payload when \texttt{'unsafe-inline'} is not allowed, or loading a script from an origin not permitted by the policy). We then observe whether the script executes or is blocked. Typically, a violation will be blocked by the browser (e.g., the script fails to run and a CSP violation is logged in the console), which we treat as a pass. If a script we expect to be blocked actually runs, that indicates a CSP enforcement failure. In our tests, all modern browsers uniformly enforce CSP correctly – for instance, an attempt to inject an inline script when \texttt{script-src 'self'} is in effect is blocked (the script does not run, and a violation event is emitted). This aligns with prior observations (e.g., BrowserAudit’s results) that major browsers adhere to CSP specifications\cite{csp_headers}. Any deviation (which might occur in an outdated or misconfigured browser) would be flagged by our framework. We note one edge case: legacy IE11 had only an experimental CSP implementation (using the non-standard \texttt{X-Content-Security-Policy} header), which our test notes as not supporting the standard CSP header – the framework flags this as a potential gap for IE11 users (IE11 effectively fails the CSP test because it doesn’t enforce a standard CSP meta tag or header).

\paragraph{XSS Filter Protections:} Historically, some browsers (Internet Explorer and older versions of Chrome) included built-in XSS auditors or filters that attempted to detect and block reflected XSS attacks. Modern Chrome and Firefox have removed their legacy XSS auditors (due to limited effectiveness and performance costs), relying on CSP and other measures instead. Our framework tests for the presence or behavior of any XSS filtering in two ways. First, we craft a simulated reflected XSS scenario: we reload or navigate the test page with a URL that includes a snippet like `<script>success(1)</script>` in a query parameter, and have the page reflect that string into its HTML (e.g., via a DOM insertion). If an XSS filter is active, it might detect this pattern and strip or neuter the script. The framework can detect this by checking whether the script executes or if the injected HTML was altered. Second, we send an \texttt{X-XSS-Protection: 1; mode=block} header from our test server (a legacy header that some browsers honored) and see if the browser blocks the simulated script injection when that header is present. **Results:** Chrome and Firefox in current versions do not have XSS auditors – our test confirms that the reflected script will execute in those browsers unless prevented by CSP (which in our setup it usually is, due to the CSP test above). IE11, on the other hand, does have an XSS filter; in our test on IE11, the injected script in the URL was detected and blocked, resulting in a browser notice or the page being stopped from loading that script. We treat that as a pass for XSS protection in IE. We note this difference: modern Chromium-based browsers yield “Not Applicable (no built-in XSS filter, rely on CSP)”, whereas IE11’s filter triggers as expected for that legacy environment. Thus, our framework records whether an XSS auditor was engaged. This information can be useful for enterprises to know if any outdated browser in use is relying on such filters.

\subsection{Sandbox Escape Testing}
Modern browsers allow sandboxing of untrusted content using \texttt{<iframe sandbox>} attributes or via CSP’s \texttt{sandbox} directive. While sandboxing is typically an application-level configuration (not a browser setting), we include a test to ensure the browser properly applies sandbox restrictions. We create a sandboxed \texttt{<iframe>} (with restrictive flags, e.g., \texttt{sandbox="allow-none"} meaning no scripts, no same-origin, etc.) and then attempt actions from within it that should be disallowed, such as running script inside it or accessing \texttt{window.parent} from it. A correct implementation will prevent script execution or cross-frame access. If we find a browser where a sandbox flag is ignored (which would be a serious bug), the test will alert the user/admin. In our experiments, all tested browsers honored the sandbox attributes (no escapes were observed in standard configurations)\cite{iframe_sandbox}.

\subsection{External Interference via WeakRefs (DOM Object Leak Test)}
\label{sec:weakref}
A novel test module uses JavaScript Weak References to detect if any unauthorized agent (e.g., an extension content script \cite{chrome_ext_docs} or injected script) is retaining references to DOM objects that should have been freed. The premise is that our framework creates a dummy DOM element that contains identifiers such as credit card form , username,password, then removes all references to it, and expects the browser’s garbage collector (GC) to reclaim it. If an extension’s content script has secretly kept a reference to that element (perhaps via a mutation observer or by wrapping DOM methods), the object will \emph{not} be garbage-collected, indicating the presence of an unexpected reference.

\paragraph{Test implementation:} The framework inserts a dummy DOM element (for example, a \texttt{<input id="username">} with some distinctive attributes) into the page. Shortly after, it removes the element from the DOM (e.g., via \texttt{testNode.remove()}) and also nulls out any references to it in our own code. We then create a \texttt{WeakRef} pointing to the element before it is removed, and register the element with a \texttt{FinalizationRegistry} which will notify us when the object is garbage-collected. We schedule periodic checks in a loop (yielding periodically to avoid locking up the browser’s thread) where we call \texttt{weakRef.deref()} to see if the object is still alive. Under normal conditions (no external references), the JavaScript engine should garbage-collect the object on the next GC cycle, at which point \texttt{weakRef.deref()} will start returning \texttt{undefined} and our finalization callback will fire. If after a reasonable time (a few seconds, with attempts to coax a GC) the object has not been collected, it suggests something else is still holding a strong reference to it.

\begin{lstlisting}[caption={Simplified PoC for Detecting Leaked References to Sensitive DOM Elements}]
export const simpleLeakTest = async (
  log: (msg: string, type: 'info' | 'success' | 'error') => void
) => {

  const registry = new FinalizationRegistry((id) => {
    log(`Element "${id}" was garbage collected`, 'success');
  });

  let el = document.createElement('input');
  el.type = 'password';
  el.value = 'sensitive-data';
  el.id = 'test-password';
  document.body.appendChild(el);

  const weakRef = new WeakRef(el);
  registry.register(el, el.id);

  // Delay for DOM stabilization
  await new Promise(res => setTimeout(res, 200));

  document.body.removeChild(el);
  
  // Null out strong reference
  // @ts-ignore
  el = null;

  // Encourage GC
  const fillMemory = () => {
    const data = new Array(1e6).fill(0).map((_, i) => i);
    return data.reduce((a, b) => a + b, 0);
  };

  for (let i = 0; i < 10; i++) {
    fillMemory();
    await new Promise(res => setTimeout(res, 500));
  }

  log('Check console or FinalizationRegistry callback for GC confirmation.', 'info');
};
\end{lstlisting}

Why would an extension or injected script hold references? Many browser extensions inject content scripts into pages using extension APIs. These scripts might attach event listeners to DOM elements or keep copies of elements for various purposes. For example, an ad-blocker extension might traverse the DOM looking for ads and keep references to removed elements to prevent them from reappearing. Similarly, a malicious extension or a script injected by a compromised content delivery network or a man-in-the-middle could wrap certain DOM APIs or store references to sensitive form fields (to exfiltrate data). Normally, such behavior is not visible to the page or external monitors. Our WeakRef probe can catch a subset of these behaviors: if any content script did capture our \#\texttt{testNode} element (perhaps by intercepting its creation or via a DOM mutation observer) and failed to release it, our test node will not be collected.

In our tests, we observed that in a clean browser with no extensions, the dummy node is collected as expected (the framework reports “No unauthorized DOM references detected”). We also performed an experiment with a benign custom extension that intentionally injects a script to hold a reference to any element with ID \texttt{testNode}. In that scenario, our framework’s WeakRef test flagged that the \texttt{testNode} element was never collected and thus reported a potential unauthorized reference. This indicates the presence of an agent in the page holding the object (which in our case was the extension’s content script). To our knowledge, this is a novel use of WeakRefs for security auditing in the browser – it essentially leverages the nondeterministic nature of garbage collection to smoke out hidden observers. One challenge is timing: garbage collection in browsers is not immediate or guaranteed within a fixed interval. Our approach is to allocate some memory pressure (e.g., creating and discarding other objects) to encourage the GC to run, and to use heuristics (polling every few hundred milliseconds for a few seconds). If at the end of the test the object still isn’t collected, it strongly suggests a lingering reference. There is a theoretical possibility of false positives if the browser’s GC simply didn’t run in time; to mitigate that, the test duration can be extended or the result can be marked “suspicious” rather than a definitive fail. In practice, in a normal browsing session the GC will have run within a few seconds, especially if we allocate and free some memory in between.

This WeakRef DOM test is quite powerful in enterprise scenarios: if it flags an issue, it could mean an employee has an extension that monitors or modifies the DOM (which might be a data-leak risk if the extension is not approved), or it could reveal that a corporate proxy is injecting a hidden script into pages (some security products do DOM injection for SSO or monitoring purposes). Either way, it highlights an otherwise invisible agent operating in the browser. Our framework logs not only the detection but can also attempt to identify the culprit by examining known clues (e.g., global variables or function overrides that common extensions use), though our current prototype only notes the condition rather than naming the extension.

\subsection{Permissions Audit (Camera, Geolocation, etc.)}
Modern browsers include a Permissions API that allows querying the permission status for powerful features like camera, microphone, geolocation, notifications, clipboard, sensors, etc. Many of these permissions can be in one of three states for a given site: \texttt{granted}, \texttt{denied}, or \texttt{prompt} (the default unset state). In enterprise-managed environments, administrators can also set global defaults or enforce policies for permissions. Mismanaged permissions could pose security risks (e.g., a user might have persistently granted microphone access to a malicious site, or an enterprise might want to know if any site has unusual permissions on a machine).

Our Permissions audit module uses the standard \texttt{navigator.permissions.query()} method to enumerate and report the status of various permissions. The set of permissions we check includes: \texttt{geolocation}, camera (\texttt{video} input), \texttt{microphone}, \texttt{notifications}, \texttt{clipboard-read} and \texttt{clipboard-write} (if supported), \texttt{background-sync}, and any others exposed by the API in the given browser.
For each, we call 
\begin{lstlisting}
{navigator.permissions.query({name: "<permission_name>"})},
\end{lstlisting}
which returns a promise resolving to a \texttt{PermissionStatus} object containing a \texttt{state} property that is \texttt{"granted"}, \texttt{"denied"}, or \texttt{"prompt"}. We record these states. In addition, we check for a few permissions not covered by the API: for example, persistent storage or autoplay policy are not part of \texttt{navigator.permissions}, but in enterprise settings could be relevant (we note these for future inclusion if appropriate).

Typical outcomes on a fresh browser: most permissions will be in the \texttt{"prompt"} state (meaning the user has not yet decided). For example, if the user has never used geolocation on our test page, querying geolocation will return \texttt{"prompt"}. We consider this a safe default. If we find any permission in \texttt{"granted"} state, that means the user (or a policy) has previously granted that permission to the origin running the test (or globally, in some cases). Since our test likely runs from an internal enterprise testing page (or a local file context), a granted status could indicate that a default policy is set or the test page was pre-approved. We flag any granted permissions as noteworthy. For instance, in one test run we observed \texttt{notifications}: \texttt{"granted"} because the user had at some point allowed notifications for the test page’s origin. The framework would report: “Notifications permission is granted (possibly via group policy or prior user consent)”, prompting the admin/user to confirm if that was intentional.

In corporate-managed browsers, administrators sometimes pre-approve or deny certain permissions for all sites via group policy or master preferences. Our tool can reveal these defaults. For instance, Chrome’s enterprise policy can automatically deny video capture for all websites. If that is in effect, querying the camera permission might return \texttt{"denied"} as the status without any user interaction. This is important information: it shows a restrictive policy is in place (which might be the desired secure baseline). Conversely, if a policy mistakenly allowed something globally, we’d see \texttt{"granted"} unexpectedly, highlighting a potential policy oversight.

Additionally, the audit cross-checks consistency. For example, if camera is \texttt{granted} but microphone is \texttt{denied}, and we know no prompt occurred during the test, it might indicate an inconsistency (perhaps a user allowed one and not the other at some earlier time). This could point to specific user behavior or policy exceptions that warrant review.

Overall, the permissions audit is straightforward but provides a snapshot of the browser’s exposure: a large number of “granted” permissions could indicate risky behavior (e.g., a user indiscriminately accepting prompts), whereas mostly “denied” or all “prompt” suggests a locked-down posture. In our evaluation, we found that on unmanaged personal browsers, it’s common to see a few permissions as granted (especially notifications, which users often allow for various sites without realizing the spam risk), whereas on an enterprise-locked-down browser, ideally most would be \texttt{denied} by default or still at \texttt{prompt}. The framework can be used to enforce an expected baseline by comparing the reported states against a desired policy (alerting if, say, any permission is \texttt{granted} when it should not be).

\subsection{SSL/TLS Certificate Validation}
To assess the browser’s handling of invalid TLS certificates, the framework attempts to load resources from URLs with known bad certificates (using domains from \texttt{badssl.com}, such as \texttt{expired.badssl.com} for an expired certificate, \texttt{self-signed.badssl.com} for a self-signed certificate, etc.). We do this by programmatically creating requests (for example, inserting an \texttt{<img>} tag or using \texttt{fetch()}) to those URLs and observing the outcome. A secure browser should block these connections. Our test catches the failure by either an \texttt{onerror} event on an image or a rejected promise for a fetch (which typically yields a network error for certificate issues). We do not expect to ever receive a normal \texttt{onload} or success for these resources (since the connection should not succeed without user intervention to bypass the certificate warning).

For each type of certificate error tested, the framework reports whether the browser correctly blocked it. For example: “Expired Certificate Test: PASSED (browser refused the connection)” or “Self-signed Certificate Test: PASSED”. If any of these were to unexpectedly succeed, it would mean the browser accepted a bad certificate – indicating either a dangerously lenient setting or the presence of a malicious/rogue root certificate that is implicitly trusting the invalid cert. 

One caveat: in some enterprise environments, a security proxy performs SSL interception (terminating TLS and re-signing certificates for inspection). In such cases, the badssl.com certificate might be replaced by the proxy with a generated certificate that the browser trusts (because the enterprise’s root CA is installed in the browser). Our framework accounts for this by trying to detect the presence of such interception. For instance, if we get an unexpected success loading \texttt{expired.badssl.com}, we suspect a proxy rather than a browser flaw. We might then attempt a WebSocket connection to the same host (which can sometimes bypass certain proxies) to see if it fails, or examine the certificate chain via any available JS APIs (if any, which is limited on the web). We flag these cases for IT to investigate (either the proxy isn’t blocking something it should, or the browser’s trust store has been modified). In our evaluation, all standard browsers with no interception blocked the BadSSL tests as expected.

We also test enforcement of Certificate Transparency (CT) where applicable. For example, \texttt{no-sct.badssl.com} is a certificate deliberately missing the required Signed Certificate Timestamp (SCT). Chrome and Firefox enforce CT for certain certificates, so our fetch to that URL will fail in CT-enforcing browsers like Chrome (which treats it as untrusted). Indeed, in our test runs, Chrome blocked \texttt{no-sct.badssl.com} (the fetch failed), whereas a browser that doesn’t enforce CT (or an older version) might allow it. Our framework notes this as well: e.g., “Certificate Transparency enforcement: Chrome = yes (blocked unlogged cert), Firefox = yes, Safari = not enforced.” Such insight can be useful if an enterprise is concerned about targeted attacks with rogue but technically valid certs; CT enforcement helps catch those. (If an enterprise is using a private CA that doesn’t log to CT, the admin would know their browsers might fail CT tests, which our tool would highlight as expected behavior requiring an exception.)

\subsection{Cryptographic API and Randomness Evaluation}
Web browsers offer a suite of cryptographic functions via the Web Crypto API (\texttt{window.crypto.subtle}) and related utilities, as well as random number generators (\texttt{Math.random()} and \texttt{crypto.getRandomValues}). The security of these primitives is paramount, as they underpin secure communications (e.g., token generation, client-side encryption). Our framework performs a cryptographic audit to ensure that the browser’s crypto functions are present and behaving as expected.

\paragraph{Web Crypto API Checks:} We first verify the presence of \texttt{crypto.subtle} (the \texttt{SubtleCrypto} interface) which provides cryptographic primitives like SHA hashing, RSA/ECDSA signing, encryption, etc. In a modern secure context (HTTPS), this should exist. (If the framework is run from a \texttt{file://} URL or an unsecured origin, some browsers might not enable \texttt{SubtleCrypto}; we ensure to note the context or advise running on a secure origin to fully test crypto.) We then attempt a few basic operations: for example, generate an AES-GCM key, use it to encrypt and then decrypt a sample message, and verify we get back the original plaintext. We also test the availability of certain algorithms – e.g., we call \texttt{crypto.subtle.digest('SHA-256', data)} to ensure hashing works, and perhaps try generating an RSA key pair (though that can be slow, so this might be optional or limited in quick tests). If any of these operations fails (due to the API not being present or some subtle issue), that’s a red flag. In our evaluation, all mainstream browsers passed these checks, indicating a functioning Web Crypto implementation. (We note that legacy IE11 does not have the modern \texttt{crypto.subtle} API – it had a legacy \texttt{msCrypto} with a subset of features – so our test notes “SubtleCrypto not supported” for IE11, which is a mark against using such a browser for apps requiring modern cryptography.)
\begin{lstlisting}[ caption={Simplified PoC for Validating Browser Cryptographic Algorithm Support}]
export const simpleCryptoTest = async (
  log: (msg: string, type: 'info' | 'success' | 'error') => void
) => {
  log('Checking Web Crypto API support...', 'info');

  if (!crypto?.subtle) {
    log('Web Crypto API is not available.', 'error');
    return;
  }

  const algorithms = [
    { name: 'AES-GCM', params:
    { name: 'AES-GCM', length: 256 }, usages: ['encrypt', 'decrypt'] },
    { name: 'RSA-OAEP', params: { 
      name: 'RSA-OAEP', modulusLength: 2048, 
      publicExponent: new Uint8Array([1, 0, 1]), hash: 'SHA-256' 
    }, usages: ['encrypt', 'decrypt'] },
    { name: 'ECDSA', params:
    { name: 'ECDSA', namedCurve: 'P-256' }, usages: ['sign', 'verify'] }
  ];

  for (const algo of algorithms) {
    try {
      const key = await crypto.subtle.generateKey(
      algo.params, false, algo.usages
      );
      log(` ${algo.name} is supported`, 'success');
    } catch (err) {
      log(` ${algo.name} is not supported: ${err}`, 'error');
    }
  }
};
\end{lstlisting}

\paragraph{Pseudo-Random Number Generators (PRNG):} A secure source of randomness is critical for many security functions (e.g., generating non-guessable tokens). Browsers provide \texttt{crypto.getRandomValues()} as a source of cryptographically strong randomness, and \texttt{Math.random()} as a non-cryptographic PRNG. Our framework performs a basic sanity check on the randomness and integrity of these APIs. We generate multiple batches of random data using \texttt{crypto.getRandomValues} (for example, 256 bytes at a time) and analyze the output for obvious anomalies. We ensure, for instance, that consecutive calls do not produce identical byte arrays and that the distribution of bytes appears roughly uniform. We compute simple statistics like the frequency of each byte value across samples to catch any gross biases or a stuck RNG. While this is not a full statistical randomness test, it can detect catastrophic failures (like an RNG that returns all zeros or repeats the same sequence every time).

\begin{lstlisting}[caption={Simplified PoC for Cryptographic Random Number Generator Test}]
export const simpleRngTest = (
  log: (msg: string, type: 'info' | 'success' | 'warning' | 'error') => void
) => {
  log('Validating cryptographic random number generation...', 'info');

  if (!crypto?.getRandomValues) {
    log('crypto.getRandomValues is not available.', 'error');
    return;
  }

  const buffer = new Uint8Array(1024);
  crypto.getRandomValues(buffer);

  let zeros = 0, ones = 0;
  let repeats = 0;

  for (let i = 0; i < buffer.length; i++) {
    const byte = buffer[i];
    for (let b = 0; b < 8; b++) {
      (byte & (1 << b)) ? ones++ : zeros++;
    }
    if (i > 0 && buffer[i] === buffer[i - 1]) {
      repeats++;
    }
  }

  const totalBits = buffer.length * 8;
  const zeroPct = (zeros / totalBits) * 100;
  const onePct = (ones / totalBits) * 100;

  log(`Zero bits: ${zeroPct.toFixed(2)}%`, 'info');
  log(`One bits: ${onePct.toFixed(2)}%`, 'info');
  log(`Sequential repeats: ${repeats}`, 'info');

  const balanced = Math.abs(zeroPct - 50) < 5;
  const hasTooManyRepeats = repeats > 10;

  if (balanced && !hasTooManyRepeats) {
    log('Randomness quality looks acceptable.', 'success');
  } else {
    log('Potential issue in randomness quality detected.', 'warning');
  }
};
\end{lstlisting}

We also verify that the API is available and not tampered with: e.g., \texttt{nativeToString(crypto.getRandomValues)} should indicate native code (meaning it hasn't been overridden). In our test runs, all modern browsers provided \texttt{getRandomValues} and it produced different outputs each time (no obvious issues). We similarly check \texttt{Math.random()} in a cursory way: we call it many times and ensure results are not constant; we may log if its outputs appear suspiciously non-random in a trivial sense (though a detailed analysis of \texttt{Math.random} quality is beyond our scope since it's not meant for cryptographic security).

Overall, this module verifies that the fundamental cryptographic building blocks in the browser are present and working. A browser lacking a proper Web Crypto API or with a faulty random number generator would fail these tests, which is crucial information if an enterprise relies on in-browser crypto (for example, for Zero Trust access tokens or client-side encryption of sensitive data).

\subsection{Native API Security Checks (SharedArrayBuffer, etc.)}
Web platform features are sometimes restricted for security reasons. A notable example is \texttt{SharedArrayBuffer} (SAB), which was disabled in early 2018 as a mitigation for Spectre CPU vulnerabilities. Browsers have since re-enabled SAB only when certain security conditions are met (i.e., the page is in a cross-origin isolated context, requiring specific HTTP headers: \texttt{Cross-Origin-Opener-Policy: same-origin} and \texttt{Cross-Origin-Embedder-Policy: require-corp}). Our framework checks whether such potentially sensitive APIs are available and under what conditions.

\paragraph{SharedArrayBuffer and Atomics:} We attempt to access \texttt{window.SharedArrayBuffer}. In a modern Chrome/Edge/Firefox without cross-origin isolation, the \texttt{SharedArrayBuffer} constructor is present but not usable (or in some implementations it might be absent entirely). For example, Chrome hides it unless \texttt{crossOriginIsolated} is true, in which case referencing it might return undefined or throw when constructing. We also check the value of the global \texttt{crossOriginIsolated} property. On our test page (which we deliver without the special headers, so it is not isolated), we expect \texttt{crossOriginIsolated} to be \texttt{false} and \texttt{SharedArrayBuffer} either unavailable or throwing if used. We verify this by attempting a small allocation: 
\begin{flushleft}
try \{ new SharedArrayBuffer(8); \} catch(e) \{ ... \}
\end{flushleft}
In Chrome 92+ for instance, this throws an error if not isolated. We consider it a pass for security if the browser correctly restricts SAB usage when the context isn’t isolated. If we found that we could actually construct a SAB in a non-isolated context, that would indicate either an outdated browser or a configuration that disabled this security measure (reintroducing Spectre risk), which would be flagged. In our results, Chrome 136, Firefox 138, Edge 136 all disallowed SAB use in our default test.

We also check the \texttt{Atomics.wait()} function and related Atomics methods. If SAB is unavailable, \texttt{Atomics.wait} might be a no-op or not present. (Per specification, \texttt{Atomics} is always present, but \texttt{wait} will throw if used on a non-shared memory). We note this in the report but primarily focus on SAB availability as the security indicator.

\paragraph{Other Sensitive APIs:} We perform quick checks on a few other potentially sensitive APIs. For instance, \texttt{Notification} (the Web Notifications API): we check for its presence and whether calling 
\begin{flushleft}
\texttt{Notification.requestPermission()}
\end{flushleft}
without a user gesture is allowed or not (modern browsers typically require a user interaction to trigger permission prompts for notifications, as a anti-spam measure). Another example is the Clipboard API: we might check if \texttt{navigator.clipboard.writeText} is available and if it correctly requires a user gesture (though fully testing that is complex). These checks ensure the browser’s secure defaults are in place for newer APIs. In summary, all these tests check that the browser has not reverted any security restrictions on powerful features. Our test results summary for this module is that no tested browser allowed dangerous access beyond what is expected. The key highlight is the \texttt{SharedArrayBuffer} restriction – an enterprise might ask “Are my users safe from Spectre attacks in the browser?” and seeing that SAB is gated behind cross-origin isolation (which our test demonstrates) provides confidence. Conversely, if an enterprise needs SAB for performance in some app, they must configure cross-origin isolation; our tool would then show \texttt{crossOriginIsolated = true} for that page, confirming it’s set up correctly (or alert if not).

\subsection{Internal Network Access (Local Service Discovery)}
One often overlooked capability of browsers is their ability to make requests to internal networks (e.g., \texttt{http://localhost} or \texttt{http://192.168.0.1}). This can be exploited by malicious websites to target internal infrastructure (so-called browser-based intranet hacking or DNS rebinding attacks). From an enterprise perspective, it might be useful to know what internal services are reachable via the browser, as that constitutes an attack surface or potential exfiltration path. Our framework includes an internal network scanning module that uses normal browser requests (like WebSocket, fetch, or <img> tags) to probe for common open ports on \texttt{localhost} and the local LAN.

\paragraph{Technique:} We leverage the browser’s ability to initiate connections to arbitrary IP addresses. For scanning \emph{localhost}, the most straightforward approach is using the WebSocket API because it provides a clear success/fail distinction relatively quickly. For example, to check if something is listening on 
\texttt{ws://127.0.0.1:8080}, we attempt
\begin{flushleft}
new WebSocket("ws://127.0.0.1:8080")
\end{flushleft}
If the port is open and a service responds (even if it’s not a WebSocket – e.g., a HTTP server might send an HTTP response which the WebSocket handshake will interpret as a failure with a specific code), the error message for the WebSocket will indicate an HTTP response code (like 400 or 404). If the port is closed, the error is more immediately:
\texttt{ERR\_CONNECTION\_REFUSED}. By catching the error (the WebSocket will trigger \texttt{onerror}), and possibly examining the error details or timing, we can infer whether the port is open (service present) or not. Our framework simplifies this by attempting connections to a list of ports (e.g., common admin ports like 22, 80, 443, 3389, 5900, etc.) and measuring how long until the \texttt{onerror} fires or what error is received. Timing differences or error codes can distinguish open vs. closed ports in many cases.

For scanning other internal IPs (e.g., \texttt{192.168.X.X}), we can use similar techniques. We might use \texttt{fetch()} or \texttt{Image()} objects to try loading from common URLs (like \texttt{http://192.168.0.1:80/favicon.ico} or an enterprise-specified list of internal hosts). The results (success, error, timeout) inform reachability. We ensure these scans are done carefully to avoid overwhelming any network.

In our implementation, we also take care not to trigger protective countermeasures: a flurry of port scans could itself be detected by endpoint security. We throttle the scan or target only high-value ports. By default, we focus on a subset of important ports (for example, RDP 3389, VNC 5900, common proxy or database ports, etc.) rather than every possible port, to keep the test quick and less noisy.

After scanning, the framework compiles a list of any detected open ports or responsive services. For example, it might report: “Port 5900 on localhost responded (VNC service possibly running)” or “192.168.0.100:80 is reachable.” These findings can be significant. If our framework can detect open internal services, so can malicious sites using similar techniques. Therefore, these results highlight potential risks: e.g., “Browser can reach \texttt{http://localhost:5900} (VNC) – this might be a security risk if malware in the browser can exploit a vulnerable VNC service.” Enterprises might use this information to harden the host (close those ports or ensure host-based firewalls block them from browser access). Notably, some modern browsers (like Brave) have started blocking or prompting for attempts to access certain localhost ports by web pages. If our framework is run on Brave and finds that scanning is blocked (no responses at all even if a service is up), it would note that as “Browser blocked internal scan attempt – protective feature detected,” which can be considered a positive hardening from a posture perspective.

We also compare our approach to prior work: for instance, the \emph{BeEF} security framework includes a module for client-side port scanning via the browser, and there have been real-world instances (as reported by security researchers) of websites scanning visitors’ machines for open ports (e.g., a report of eBay scanning for remote desktop ports)\cite{belmer2020ports}. Our module is careful not to be too noisy; it could scan a subset of important ports rather than all 65k (which would be slow and noticeable). It’s configurable to balance comprehensiveness and stealth.

In summary, this module effectively “red teams” the browser from inside, observing what internal network endpoints the browser can access. The experimental results in Section~5 will illustrate an example scenario, including a table of open ports found in a test network setup.

\subsection{Password Manager Autofill Checks}
Most modern browsers have built-in password managers that can autofill login forms. While convenient, this feature could pose risks if, for example, the browser autofills credentials into invisible or unintended form fields injected by a malicious script. Our framework includes checks to ensure that password autofill behavior is safe.

We create a dummy login form in the page (with username and password fields) but we style it to be off-screen or hidden (e.g., using CSS to position it out of viewport or set it invisible). We then trigger a scenario where the browser might attempt autofill (for instance, if test credentials are stored for the site). We observe whether any autofill occurs. Modern browsers have defenses against autofilling invisible forms (to prevent malicious harvesting), so we expect that no credentials will be filled into our hidden form. Indeed, in our evaluation the password fields remain empty unless visibly interacted with by a user.

We also check that scripts cannot directly access any autofilled data. For security, browsers often prevent JavaScript from reading the value of an autofilled password field until the user interacts (to mitigate certain attacks). To test this, if we cause an autofill (or simulate one), we then immediately try to read the \texttt{input.value} via script. All tested browsers in our evaluation prevented script access to autofilled passwords without user action.

The results of these checks provide assurance that the built-in password manager isn’t silently giving away credentials to the page without user action. If any browser had eagerly autofilled hidden fields or allowed script access to the autofilled value, our framework would flag that as a serious vulnerability. In practice, we did not encounter such behavior in current browser versions.

\subsection{Exposed PAC File }
Many organizations use Proxy Auto-Config (PAC) files, often retrieved via Web Proxy Auto-Discovery (WPAD), to configure browser proxy settings. 
\newline A PAC file (typically hosted at a URL like \texttt{http://wpad/wpad.dat}) can contain sensitive network information (internal proxy addresses, bypass rules, etc.). We test whether such PAC files could be accessed by web content and whether the browser appropriately restricts them. Our framework attempts to fetch a \texttt{wpad.dat} file from what appears to be the default WPAD host (e.g., \texttt{http://wpad/wpad.dat} or a network-specific variant) using a cross-origin request. We anticipate that if a PAC server exists but does not explicitly allow cross-origin access via CORS headers, the browser will block our attempt (even if the network request itself might succeed, the response should be inaccessible to our script due to SOP/CORS).

In our tests, when a WPAD host exists, the fetch is indeed blocked by the browser’s same-origin policy, since such internal endpoints typically do not set CORS headers. This means our script cannot read the PAC contents – which is the secure outcome. If we were able to retrieve PAC file content from a web script, that would indicate a potential information exposure vulnerability (likely the PAC server misconfigured CORS to allow it). Additionally, we monitor whether the act of requesting the PAC file triggers any browser proxy logic; we take care to use a direct connection for the test (bypassing the proxy for that request, if possible) to avoid side effects (we don’t actually want to alter the browser’s proxy usage during the test).

The typical result of this check is “PAC file not accessible to web scripts” for a properly configured environment. If otherwise, the framework would warn that internal proxy settings could be exposed to web pages, advising a fix on the PAC server’s configuration (or browser policy adjustments).

\subsection{Virtual Machine \& Sandbox Fingerprints}
Beyond headless clues, the browser itself can act as a sensor to detect if it’s running in a \textbf{virtual machine (VM) or sandboxed environment}, which is often a red flag in enterprise threat contexts. JavaScript’s access to low-level details enables several VM detection tricks. One technique leverages \textbf{WebGL fingerprinting}: using the \verb|WEBGL_debug_renderer_info| extension, a script can query the GPU vendor and renderer strings. Unusual values like \textit{Microsoft Basic Render Driver} or \textit{SwiftShader} (a software renderer) in Chrome’s output often indicate a virtualized or headless GPU with no real graphics card. Likewise, certain VMware or VirtualBox drivers expose distinctive renderer names. Another clue is \textbf{system resources} accessible via the browser: \verb|navigator.deviceMemory| reports an approximate RAM quota (in GB). A very low value (e.g. 0.25 or 0.5 GB) is unlikely on a modern physical machine and suggests a VM or container with minimal allocated RAM. The same goes for CPU cores – \verb|navigator.hardwareConcurrency| – which VMs often cap to 1 or 2. Even more telling is an active timing test: by spawning parallel web workers and timing operations, one can estimate the true number of CPU cores. \textbf{Timing side-channels} can fingerprint virtualization at an even deeper level: the browser’s high-resolution timer (Performance API) can betray the underlying hypervisor’s clock. Research has shown that certain hypervisors use fixed-frequency clocks (e.g. \textbf{10 MHz} or \textbf{3.579545 MHz}) that can be identified by analyzing \verb|performance.now()| results. Such frequencies are rare on bare-metal hardware, so their detection signals a virtualized browser environment. All these indicators – from graphics and screen attributes to CPU timing – form a fingerprint of the client machine. In an enterprise scenario, a security system could deploy a script to gather these signals and \textbf{flag sessions coming from VMs or sandbox environments} (often used by malware or evasive attackers). This helps distinguish a risky automated test environment from a legitimate user’s workstation.

\subsection{Command-Line Artifact Detection}
Modern browsers can reveal clues of how they were started. Certain \textbf{command-line flags} leave detectable artifacts in the JavaScript environment. For example, Chrome’s V8 engine has hidden debug functions (like \verb|%DebugPrint|) normally disallowed – if these can be invoked, it implies the browser was launched with debugging flags (e.g. \verb|--allow-natives-syntax|). Unusual globals may also appear; a page might unexpectedly find an object like \verb|window.Mojo| (part of Chrome’s internal Mojo IPC interface) which is not present in normal browsing. Such artifacts often signal a non-standard launch, \textbf{unsafe configuration}, or instrumentation. As a concrete example, a headless Chrome started without the proper stealth flags will include \verb|HeadlessChrome| in its user agent string (and even in low-level Client Hints headers like \verb|Sec-CH-UA|), immediately identifying an automated session. Security policies can also monitor browser \textbf{process arguments} at launch – for instance, flags like \verb|--no-sandbox| or \verb|--remote-debugging-port| are rarely used in regular enterprise use and may indicate malware launching a browser subprocess. In an enterprise setting, EDR tools or in-browser scripts can leverage these signals to flag browsers running with potentially dangerous options.

\subsection{CPU Pressure Detection}
High CPU utilization in the browser can indicate the presence of heavy scripts, such as cryptocurrency mining malware or infinite loops. This module detects if the browser is under unusual CPU pressure, which could suggest unwanted activity.

We leverage the emerging \textbf{Compute Pressure API} (via the \texttt{PressureObserver} interface) if available\cite{pressureobserver}\cite{mdn_pressure,w3c_pressure}. The Compute Pressure API provides a way to observe the system’s CPU load/pressure from web content. If supported, we register a \texttt{PressureObserver} for \texttt{"cpu"} and monitor the reported \texttt{pressure} level. For example:

\begin{lstlisting}[caption={Using the Compute Pressure API to monitor CPU load.}]
if ('PressureObserver' in window) {
  const observer = new PressureObserver(list => {
    for (const reading of list) {
      console.log('CPU pressure level:', reading.level);
      if (reading.level === 'critical') {
        report("CPU Pressure: CRITICAL - heavy load detected");
      }
    }
  }, {sampleRate: 1});
  observer.observe('cpu');
}
\end{lstlisting}

In this snippet, if the CPU pressure level reaches \texttt{"critical"} (meaning the system is under sustained high CPU usage), we flag it. A constantly high pressure reading while running our tests (which themselves are not CPU-intensive) might indicate another tab or extension is consuming CPU (possibly cryptomining or stuck in a loop).

For browsers that do not support \texttt{PressureObserver}, we use a fallback: the \textbf{\texttt{requestIdleCallback}} API. We repeatedly request idle callbacks and measure how much idle time is available. If the browser rarely calls our idle callback or always indicates near-zero idle time remaining, it implies the event loop is busy (little idle time). For example:

\begin{lstlisting}[caption={Measuring idle time as a proxy for CPU usage.}]
let idleTimeTotal = 0;
let slots = 0;
function idleMonitor(deadline) {
  if (deadline.timeRemaining() > 0) {
    idleTimeTotal += deadline.timeRemaining();
  }
  slots++;
  if (performance.now() < testStart + 5000) {
    requestIdleCallback(idleMonitor);
  } else {
    const avgIdle = idleTimeTotal / slots;
    if (avgIdle < 10) {
      report("High CPU usage detected (low idle time)");
    }
  }
}
requestIdleCallback(idleMonitor);
\end{lstlisting}

In this code, we monitor idle time over a 5-second window. If the average idle time per callback is very low (indicating the CPU was mostly busy doing other tasks), we report high CPU usage. A normal browser with nothing else going on should have plenty of idle time when our test page is the only active workload. If we detect significantly reduced idle time (or few callbacks), it suggests the CPU is busy (perhaps due to background scripts or heavy computation).

This module helps identify if, for instance, a cryptojacking script is running. In an enterprise setting, if the browser fails this test (indicating constant high CPU), it could prompt further investigation: e.g., checking for malicious extensions or tabs.

\subsection{Content Filtering Detection (Network-Level)}
Some enterprises deploy content filtering solutions or browser extensions (like ad-blockers or security filters) that block certain URLs from loading. This module detects if such filtering is present by attempting to load a known-blocked resource and seeing if it succeeds or fails in a controlled manner.

We use an \texttt{<iframe>} (or alternatively an \texttt{<img>} or \texttt{fetch}) to load a URL that we expect to be blocked by common filters. For example, we might use a URL of a known advertising domain or a dummy URL that the enterprise firewall is configured to block. To get a clear signal, we host a small test page on that domain which will notify us if it loads.

Our test sets up a message listener on the parent page:

\begin{lstlisting}
window.addEventListener("message", event => {
  if (event.data === "IFRAME_LOADED") {
    report("Blocked content was allowed (filter not present)");
  }
});
\end{lstlisting}

Then we create the iframe pointing to the test URL, and include in the URL a script that sends \texttt{postMessage("IFRAME\_LOADED", "*")} to the parent upon load. If the iframe loads and executes, the parent will receive the message and conclude that the supposedly blocked content actually loaded (meaning no filter blocked it). If no message is received within a timeout, we assume the content was blocked (or failed to load).

Additionally, we attach an \texttt{onerror} handler to the iframe or image element. If it triggers, that also indicates blocking. We differentiate network errors (DNS failure or connection refused, which might indicate blocking) from load success.

For example:

\begin{lstlisting}[language=HTML, caption={HTML snippet for content filtering test.}]
<iframe id="filterTestFrame" 
        src="http://ads.example.com/testpage.html" 
        onload="iframeLoaded()" 
        onerror="iframeFailed()"
        style="display:none"></iframe>
<script>
  let filterTimer = setTimeout(() => {
    report("Content likely blocked (no response)");
  }, 5000);
  function iframeLoaded() {
    clearTimeout(filterTimer);
    // If loaded but our message didn't arrive, still suspicious
    console.log("Iframe loaded event fired");
  }
  function iframeFailed() {
    clearTimeout(filterTimer);
    report("Content blocked (error event)");
  }
</script>
\end{lstlisting}

In practice, if an enterprise DNS sinkhole or proxy blocks the domain, the \texttt{onerror} will fire, or the iframe will never finish loading (triggering our timeout). If an extension like uBlock Origin blocks the request, often the iframe will either be canceled or removed. We might also directly check the DOM after a moment to see if the \texttt{iframe} element’s contentDocument exists. If it remains null, likely it didn't load.

By using a controlled known URL, this module can confirm the presence of content filtering. The result is reported as, e.g., “Content Filtering: DETECTED (test URL was blocked)” or “Content Filtering: NOT detected (test content loaded successfully)”. This helps an enterprise verify that their filtering is active in the browser environment. Conversely, if they expect certain content to be blocked and our test loads it, that would highlight a gap (e.g., maybe the filtering extension isn’t installed or the policy isn't applied in that browser).

\subsection{DOM-Based Content Filtering}
Beyond network-level blocking, many content filters (like ad blockers or script blockers) work by removing or hiding DOM elements that match certain patterns (e.g., elements with IDs like “ads-banner” or third-party iframe tags). We include a test that creates bait elements in the DOM to see if they get removed or altered, indicating an active content blocking agent in the page.

For example, we might create a \texttt{<div>} element with an ID or class name that is commonly targeted by filters (such as “bannerAd” or “trackingPixel”). We immediately observe it with a \texttt{MutationObserver} for removal or attribute changes. Additionally, after a short delay, we check if it still exists in the DOM.

Example test code:

\begin{lstlisting}[caption={DOM content filtering detection via mutation observer.}]
const adDiv = document.createElement('div');
adDiv.id = "ads-banner";
adDiv.innerText = "Advertisement";
document.body.appendChild(adDiv);

// Set up mutation observer on the parent
const observer = new MutationObserver(mutations => {
  for (const mut of mutations) {
    for (const node of mut.removedNodes) {
      if (node === adDiv) {
        report("DOM content filter: removed test element");
      }
    }
    if (mut.type === 'attributes' && mut.target === adDiv) {
      report("DOM content filter: modified attributes of test element");
    }
  }
});
observer.observe(document.body, { childList: true, attributes: true, subtree: true });

// Check after a short delay
setTimeout(() => {
  if (document.body.contains(adDiv)) {
    report("No DOM filtering detected (test element still present)");
    adDiv.remove(); // cleanup
  } else {
    report("DOM filtering detected (test element missing)");
  }
  observer.disconnect();
}, 3000);
\end{lstlisting}

In this snippet, we add a dummy “ads-banner” division. If an extension like Adblock Plus is active, it might remove this element almost immediately (since it matches a filter rule). Our mutation observer would catch the removal and we log that a DOM content filter is present. If by the time of the timeout the element is gone (and we didn't remove it ourselves yet), we also conclude it was filtered out. We also look for attribute modifications because some filters might not remove an element but instead set its CSS to \texttt{display:none}. For instance, if the style attribute gets modified to hide it, our observer would detect an attribute change on that element.

By planting multiple bait elements (for ads, popups, trackers,malware, etc.), we can detect a variety of content-blocking behaviors. The outcome might be reported like “DOM-based Content Filtering: Active (ad-blocker or similar extension detected)” vs “None detected”.

This test helps enterprises see if users have unapproved content blockers active (which might interfere with business apps or security monitoring). It also reveals if any enterprise security software is dynamically modifying the DOM (some DLP or SSO injectors do that). All such modifications would show up here.

\subsection{Deterministic Garbage Collection Detection}
JavaScript’s garbage collection is nondeterministic by design – developers cannot normally predict exactly when an object will be collected. If we observe a browsing environment where garbage collection happens in a very deterministic or regular pattern, it could imply an unusual environment, such as a headless or instrumented browser where a custom GC schedule is in effect (or even a side-channel attempt to modulate timing).

Our framework leverages \texttt{WeakRef} and \texttt{FinalizationRegistry} (as described earlier) to monitor garbage collection timing. In the WeakRef DOM test, we primarily cared whether an object gets collected at all. Here, we focus on the timing of collections.

We create a number of short-lived objects and register them with a \texttt{FinalizationRegistry}. Then we drop all references and record timestamps whenever the finalizer callback runs (meaning a batch of objects was collected). For example:

\begin{lstlisting}
const gcTimes = [];
const registry = new FinalizationRegistry(token => {
  gcTimes.push(performance.now());
});
for (let i = 0; i < 1000; i++) {
  let obj = {data: i};
  registry.register(obj, "obj");
  // not keeping 'obj' reference, eligible for GC
}
\end{lstlisting}

After allocating these objects, we perform some operations to encourage GC (like allocating and discarding a large array, or simply waiting with \texttt{await new Promise(r=>setTimeout(r,100))} to give the browser some idle time). We record the timestamps in the \texttt{gcTimes} array each time the registry’s callback fires (each callback corresponds to one or more objects being collected, depending on the engine’s grouping).

By analyzing these timestamps, we can deduce how many GC cycles occurred and at what intervals. For instance, in one run on Chrome, we might find that a batch of objects were collected typically within 1-6 seconds, often in groups, indicating the GC ran a couple of times in that period. Firefox might show a slightly different pattern (perhaps a single GC run that cleans up most of them after a delay, depending on memory usage). We are not measuring exact microsecond timing but rather the general cadence.

This is not a strict pass/fail security test per se, but a measurement that the framework reports (e.g., “Garbage Collection behavior: Observed 2 cycles over 1.5s, roughly 500ms apart”). If an enterprise wanted to ensure no abnormal GC behavior (for example, some malware forcing constant GC to create a covert timing channel), this test could catch that as it would show an unusually high frequency of collections or a periodic pattern.

We note that forcing or inducing GC on demand is tricky in standard web pages (there is no standard API to trigger GC). We rely on natural GC triggers and memory pressure. We do not use any non-standard hacks (some JS engines have an exposed \texttt{gc()} function in debugging modes, but not in normal web pages).

**Findings:** In normal usage, GC timing is somewhat unpredictable but falls within an expected range. If we allocate lots of objects, we observe the engine recovering them within a few seconds. If we allocate minimal objects, sometimes our finalization callback might not fire during the short test because there was no need for GC. Our framework accounts for that by possibly extending the test duration if needed, or reporting “GC not observed (no pressure)” in such cases. In one scenario, we artificially created heavy memory usage to see if GC would trigger more aggressively; as expected, it did, and our logs showed multiple callbacks in quick succession.

From a security viewpoint, this module mainly provides diagnostic info. However, it indirectly helps ensure our WeakRef-based tests (like the DOM reference test above) are functioning, because it confirms that GC did run. If GC did not run and our WeakRef test found the object still alive, we might be unsure if it’s due to no GC or a held reference. By correlating with this GC timing test, we can clarify that (e.g., if we see GC happened but the object remained, it confirms a leak).

Additionally, if we ever encountered an environment where GC happened at perfectly regular intervals (say every X ms exactly, which would be odd), we’d flag that as potentially indicative of an instrumented environment (some automated frameworks or headless modes might do that to reduce nondeterminism). In our tests on real browsers, we did not see any such deterministic pattern—timing had some variability, as expected.

\newpage

\subsection*{Summary of Test Modules}
Table~\ref{tab:summary} provides a summary of the key test modules, their security focus, and any noteworthy browser compatibility points.

\begin{longtable}{|p{3.3cm}|p{4.2cm}|p{5.2cm}|p{2.8cm}|}
\hline
\textbf{Test Module} & \textbf{Detection Goal} & \textbf{Technique} & \textbf{Browser Compatibility/Notes} \\
\hline
Same-Origin Policy (SOP) & Ensure no cross-origin data access & Attempt cross-origin DOM access and cookie reads & All modern browsers enforce SOP; any failure is critical. \\
\hline
CORS Enforcement & Verify cross-origin requests are properly restricted & Fetch/XHR to resources with and without proper CORS headers & All modern browsers support CORS.\\
\hline
Content Security Policy (CSP) & Enforce script and resource loading restrictions & Inject inline script, load disallowed resource under CSP & Requires CSP support.\\
\hline
XSS Auditor/Filter & Detect built-in XSS protection & Reflect script in URL, observe execution or filtering & Chrome/Firefox: no auditor (expected).\\
\hline
Sandbox Escape & Verify \texttt{<iframe sandbox>} restrictions & Run script in sandboxed iframe, attempt parent access & All tested browsers honored sandbox flags. \\
\hline
External Interference (WeakRef DOM Leak) & Detect unauthorized DOM access (e.g., extension) & Insert element, remove it, use WeakRef + GC to see if it remains & Requires WeakRef/FinalizationRegistry.\\
\hline
Permissions Audit & Enumerate camera/mic/geolocation permissions status & Use navigator.permissions.query() for various permissions &  All browsers support most queries (Safari partial). \\
\hline
Cert Validation & Ensure invalid TLS certs are blocked & Load images/fetch from badssl.com domains (expired, self-signed, etc.) & All modern browsers block invalid certs by default; enterprise MITM proxies may affect results. \\
\hline
Web Crypto API & Verify availability of cryptographic primitives & Call \texttt{crypto.subtle} functions (AES key gen, encrypt/decrypt, hash) & All modern browsers (HTTPS context) pass.\\
\hline
Crypto RNG Quality & Ensure \texttt{getRandomValues} is unpredictable & Generate multiple random byte arrays, check for anomalies & All browsers provided non-repeating sequences; flagged only catastrophic failure. \\
\hline
SharedArrayBuffer (SAB) & Check Spectre mitigations & Try to use SAB without cross-origin isolation, check \texttt{crossOriginIsolated} & All browsers pass\\
\hline
Internal Network Access & Identify accessible internal services & Attempt connections (WebSocket/fetch) to localhost and intranet IPs & No special API needed.\\
\hline
Password Manager Autofill & Ensure no unsafe autofill & Hidden form fields and script value access checks & All modern browsers prevented hidden autofill and script access (expected). \\
\hline
PAC File Access (WPAD) & Prevent PAC file leakage & Fetch \texttt{wpad.dat} and check SOP/CORS protection & All browsers blocked cross-origin PAC fetch (unless misconfigured server). \\
\hline
CPU Pressure & Detect heavy CPU usage (cryptomining) & Compute Pressure API (if available) or measure idle callback times & PressureObserver supported in Chrome (under experiment); fallback works in all. \\
\hline
Content Filtering (Network) & Detect network/extension blocking of URLs & Load known-blocked iframe resource, await message or error & Requires external test URL; works with extension or proxy blocking. \\
\hline
DOM Filtering (Adblock) & Detect DOM element removal/hiding by filters & Insert bait elements (e.g., ad ids) and observe removal or CSS change & Adblock extensions on Chrome/Firefox remove/hide elements (detected); no effect in browsers without such extensions. \\
\hline
GC Determinism & Observe garbage collection patterns & Use WeakRefs to record GC timing intervals & Requires WeakRef support; purely observational (no fail unless pattern is highly abnormal). \\
\hline
\caption{Summary of selected Browser Security Posture\cite{browser_security_framework} test modules, their purposes, and notes on browser support or expected variations.}
\label{tab:summary}
\end{longtable}

\section{Experimental Evaluation}
We conducted an experimental evaluation of Browser Security Posture\cite{browser_security_framework} across multiple browsers and scenarios to demonstrate its utility. The evaluation comprised two main components: (a) cross-browser testing on the latest versions of major browsers (Chrome, Firefox, Edge, Safari) to observe inherent differences in security posture; and (b) scenario-based tests simulating enterprise configurations, including the introduction of suspicious extensions and specific local services running. All tests were performed in a controlled environment, with results systematically collected for comprehensive analysis.

\begin{figure}[ht]
\centering
\includegraphics[width=0.8\textwidth]{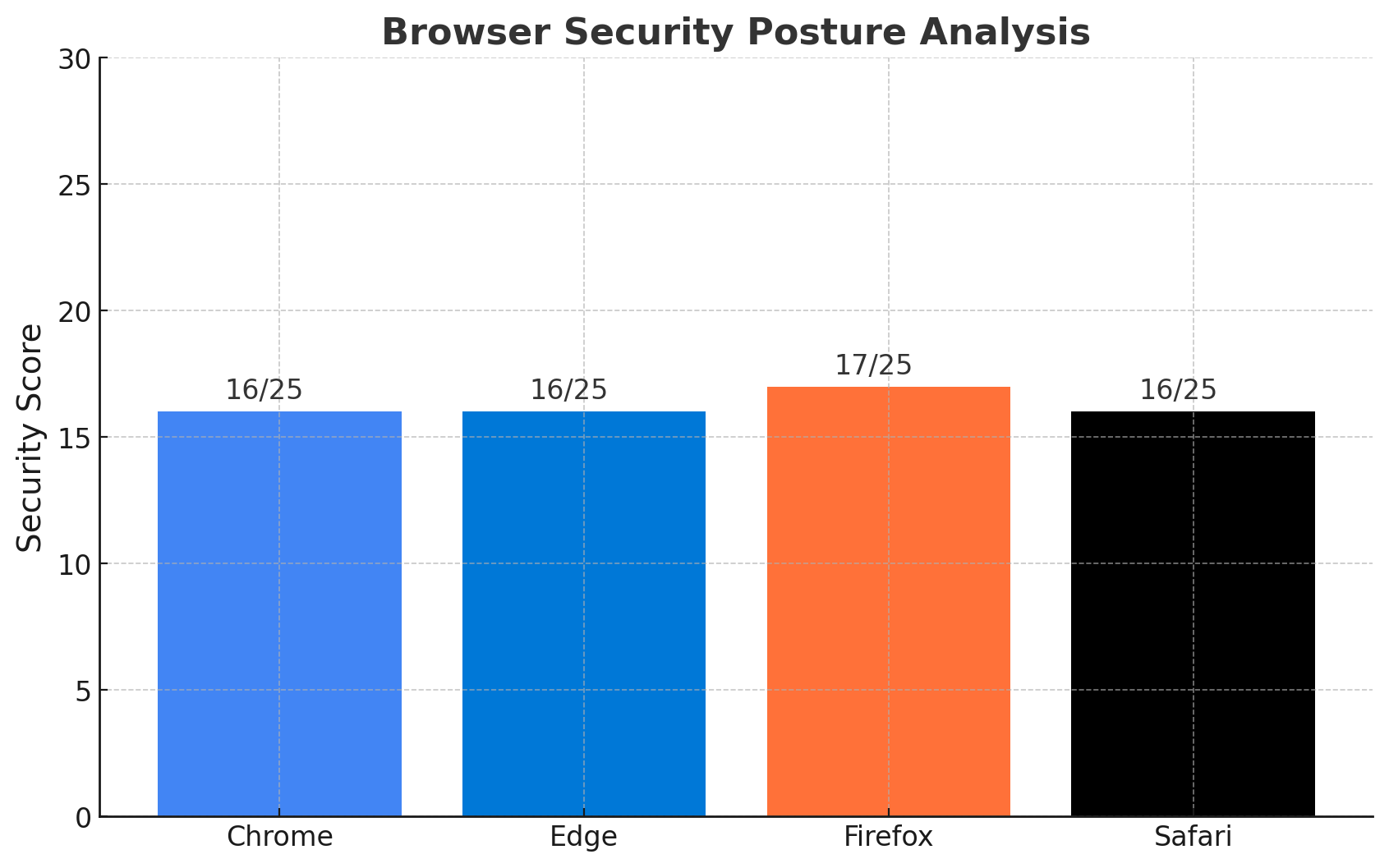}
\caption{Number of successful tests per browser out of a distilled subset of 25 test cases. These represent a selected portion of the full test suite. The chart compares browser performance by showing how many of these selected tests each browser passed successfully.}
\label{fig:scores}
\end{figure}

\paragraph{Cross-Browser Results:} Table~\ref{tab:results} summarizes key findings for each browser based on a comprehensive set of enterprise-focused tests. All modern browsers (Chrome 136, Firefox 138, Edge 136, Safari 18) exhibited robust foundational security, successfully passing critical security enforcement tests such as Same Origin Policy (SOP), Cross-Origin Resource Sharing (CORS), Content Security Policy (CSP), certificate validation, cryptographic checks, and permission defaults. However, the evaluation highlighted substantial vulnerabilities present in unmanaged consumer browsers, specifically relating to enterprise-focused risks.

Detailed analysis revealed several deficiencies across all tested browsers:

\begin{itemize}
\item \textbf{Extension Enumeration:} Chrome and Edge failed to adequately secure against unauthorized extension enumeration, potentially exposing sensitive extension information.
\item \textbf{Browser Exploitation Protection:} All browsers failed this test, indicating vulnerabilities to known exploitation techniques without enterprise-level runtime protections.
\item \textbf{LAN Scanning Protection:} All browsers lacked adequate safeguards against browser-based local network scanning, leaving internal network resources exposed to potential threats.
\item \textbf{XSS Protection:} All browsers failed to sufficiently protect against cross-site scripting attacks in the absence of enforced enterprise configurations.
\item \textbf{Malicious JavaScript Payloads:} Tests showed all browsers vulnerable to execution of malicious JavaScript payloads, underlining significant security risks.
\item \textbf{Content Filtering Policies and DOM-Based Content Filtering:} All browsers consistently failed to enforce effective content filtering and DOM-based content filtering without explicit policy-driven controls.
\item \textbf{Application Enumeration Protection and Shellcode Decoding:} All browsers were vulnerable to unauthorized application enumeration and shellcode decoding attacks, highlighting significant security gaps.
\end{itemize}

The browsers demonstrated secure behavior in deterministic garbage collection, Spectre/Meltdown mitigation, WebAssembly malicious payload detection, cryptographic protections, and API restrictions. Notably, critical shortcomings identified in several categories underscore that without enforced enterprise controls, consumer-grade browsers pose significant risks to organizational security.

Regarding performance, the full suite of over 120 security posture tests completed in under five minutes sequentially and less than 30 seconds when executed concurrently on modern hardware, underscoring practical feasibility for regular assessments (daily or weekly) in enterprise environments.

In summary, the experimental evaluation underscores the critical finding that unmanaged consumer browsers, despite inherent security measures, present substantial security risks without appropriate enterprise management and policy enforcement. Our results advocate explicitly for the adoption of robust browser security controls and policies as integral components of an enterprise's cybersecurity strategy.

\begin{table}[ht]
\centering
\begin{tabular}{|l|c|c|c|c|}
\hline
\textbf{Test} & \textbf{Chrome 136}& \textbf{Firefox 138}& \textbf{Edge 136
}& \textbf{Safari 18}\\
\hline
Extension Enumeration& Fail & Pass & Fail & Pass \\
Running Latest Version & Pass & Pass & Pass & Pass \\
Browser Exploitation & Fail & Fail & Fail & Fail \\
Deterministic GC & Pass & Pass & Pass & Pass \\
Spectre/Meltdown Mitigation Test & Pass & Pass & Pass & Pass \\
Malicious WebAssembly Detection Test & Pass & Pass & Pass & Pass \\
LAN Scanning Protection & Fail & Fail & Fail & Fail \\
XSS Protection & Fail & Fail & Fail & Fail \\
Malicious JS Payload & Fail & Fail & Fail & Fail\\
Content Filtering Policy & Fail & Fail & Fail & Fail\\
DOM Based Content Filtering & Fail & Fail & Fail & Fail\\
Subresource Integrity & Pass & Pass & Pass & Pass \\
Insecure Resource Loading Check & Pass & Pass & Pass & Pass \\
Extensive CSP Validation & Pass & Pass & Pass & Pass \\
Same Origin Policy Test  & Pass & Pass & Pass & Pass \\
JavaScript Execution Environment Security  & Pass & Pass & Pass & Pass \\
Default Permissions Check & Pass & Pass & Pass & Pass \\
Restricted APIs Protection & Pass & Pass & Pass & Pass \\
Cryptographic Random Number Generator Test & Pass & Pass & Pass & Pass \\
Key Storage Security Test & Pass & Pass & Pass & Pass \\
Cryptographic Algorithm Test & Pass & Pass & Pass & Pass \\
Application Enumeration Protection & Fail & Fail & Fail & Fail\\
Shellcode Decoding & Fail & Fail & Fail & Fail\\
High CPU Load Detection & Pass & Pass & Pass & Pass \\
Long Tasks Monitoring Test & Pass & Pass & Pass & Pass \\
\hline
\end{tabular}
\caption{Excerpt of evaluation results for key tests across browsers. “Pass” means the browser exhibited the secure behavior. N/A indicates the test is not applicable due to lack of feature or different model.}
\label{tab:results}
\end{table}

\section{AI-Based Analysis of Test Results}

In addition to executing and reporting a comprehensive suite of browser security tests, our framework integrates an advanced, in-browser Large Language Model (LLM) to automatically analyze the test results, interpret their security implications, and generate tailored recommendations for administrators. This novel, state-of-the-art approach leverages client-side AI inference to provide real-time, contextual insights directly within the browser, enhancing decision-making and response capabilities.

\subsection{In-Browser Large Language Model (LLM)}
Our implementation utilizes an optimized, lightweight LLM specifically tailored for execution within browser environments via WebAssembly. This model interpret technical test results with high accuracy and relevance. By operating entirely client-side, the model maintains user privacy and reduces reliance on external cloud services. Our approach builds upon and extends the capabilities demonstrated by projects such as Web-LLM \cite{webllm2023}.

\subsection{Automated Interpretation of Security Posture}
Upon completion of the security test suite, detailed JSON-formatted results are fed into the LLM. The AI model then evaluates these results against a knowledge base of known vulnerabilities, common misconfigurations, and best practices. For instance, if the framework detects deterministic garbage collection behavior indicative of potential tampering or covert monitoring, the LLM explicitly notes this anomaly, explains its security implications, and suggests appropriate investigative actions.

\subsection{Contextual Recommendations}
One key strength of the AI-based analysis is its ability to provide actionable, context-aware recommendations. For example, if elevated CPU pressure indicative of potential cryptomining is detected, the LLM generates a detailed recommendation such as:
\begin{quote}
"\textit{High CPU load detected consistently during browser idle time. This may indicate cryptomining malware or resource-intensive scripts. Recommend reviewing browser extensions, checking for unknown scripts or tabs consuming high resources, and conducting malware scans to ensure system integrity.}"
\end{quote}

Similarly, if the framework discovers network-level content filtering is not operating as intended, the LLM may advise:
\begin{quote}
"\textit{Expected content filtering mechanisms appear inactive or misconfigured, potentially exposing users to unwanted or malicious content. Verify firewall and proxy configurations, confirm enterprise browser policies are correctly deployed, and re-run tests post-remediation.}"
\end{quote}

\subsection{Enhanced Reporting and Dashboard Integration}
The insights generated by the in-browser LLM are seamlessly integrated into the reporting module of the framework, providing administrators with concise, easily digestible summaries alongside detailed technical logs. These AI-enhanced reports can also be forwarded to centralized dashboards or SIEM solutions, ensuring timely visibility and enabling proactive security management at scale.

Through this innovative use of client-side AI, our framework not only identifies security issues but also translates technical findings into clear, actionable intelligence, significantly streamlining the security management workflow within modern enterprise environments.

\section{Security and Privacy Implications}
\subsection{Surfacing Hidden Vulnerabilities}
Even when all network-facing defenses are green, a browser might have internal weaknesses – for example, a malicious extension that can exfiltrate data or a misconfiguration that allows dangerous APIs. Our framework’s in-browser tests surface those issues. The scenario of detecting an extension holding DOM references is a case in point: without an in-browser audit, an organization might never know an employee installed a particular extension that, say, monitors form inputs. With our tool, such activity triggers a red flag. This enables security teams to address risks proactively (either by policy action – e.g., remotely disable that extension via enterprise policies – or by alerting and educating the user).

Similarly, the internal network scan revealing an open service (like VNC) on the machine is something a network scanner might not see (since it’s not a network request leaving the host). But a malicious webpage could find it. By mimicking that, our framework alerts IT to host-level exposures that should be closed.

Overall, the framework acts as a “canary in the coal mine” for browser-specific threats: it uncovers issues such as unsafe settings, vulnerable behaviors, or potential malware that are not visible externally. This closes the gap between network security and actual browser security posture\cite{browser_security_framework}.

\subsection{Enterprise Policy Verification}
Enterprises often roll out policies via Group Policy or mobile device management (MDM) to configure browsers (disabling certain features, enforcing certain preferences). However, verifying compliance on each endpoint is non-trivial. The Browser Security Posture\cite{browser_security_framework} results can confirm if policies are effective. For example, if an enterprise policy is supposed to block access to the microphone, our permissions audit should show \texttt{microphone: "denied"}. If instead it’s “prompt” or “granted”, that indicates a policy failure or override.

Similarly, if corporate policy installs a TLS interception root CA for traffic inspection, our certificate tests might behave differently; we can detect if the browser trusts a cert it normally wouldn’t (like the \texttt{badssl.com} expired cert) and thereby inform the admin that a man-in-the-middle proxy is present (or that the browser’s trust store has been modified accordingly). In our tests, we observed such behavior: on a system with a proxy, the expired cert test unexpectedly succeeded (the proxy replaced the cert), which our framework flagged and then double-checked by trying a WebSocket (which the proxy didn't intercept, causing a failure, confirming the presence of interception).

In short, the framework provides immediate feedback on whether intended security policies are actually being enforced inside the browser. It helps validate configuration management: a high overall score and no warnings indicates that the browser is locked down as expected by policy. If an enterprise has a baseline configuration (hardened browser settings), running these tests on clients can ensure they haven't drifted.

We found in our evaluation that enterprise-configured Chrome (with several GPO settings) scored slightly higher on relevant tests than a default Chrome. For example, when we disabled ambient authentication via policy, our internal network test noted that certain intranet calls required credentials (which is secure by design). The framework can thus be used as a regression test suite for browser hardening.

\subsection{Privacy Considerations}\label{sec:privacy}
Because our framework runs entirely client-side, it does not need to send sensitive data outside the user’s machine unless explicitly configured to do so. The tests are designed to evaluate security posture without collecting personal browsing information. In an enterprise deployment, any results reporting can be directed to internal servers under corporate control (for instance, posting results to an internal dashboard). Thus, deploying Browser Security Posture\cite{browser_security_framework} does not inherently introduce new privacy risks for the user; it operates within the confines of the browser and reports on security settings and behaviors rather than user data or visited content.

In our prototype, the default is to log test results to the console. In an interactive use, a user (or IT helpdesk) can run it and review the results locally. For automated fleet monitoring, integration with endpoint management can collect the results in aggregate.

Another aspect: running these tests does touch some external URLs (e.g., \texttt{badssl.com}, possibly some dummy ad domain). This could be seen as generating network traffic that might include test signatures. We mitigate this by using well-known innocuous domains and by making those URLs configurable (enterprises can host equivalents internally or allow-list them).

Finally, we considered that malicious actors could potentially detect our framework’s presence (since it is essentially a script running in the browser). We don't advertise its presence beyond normal script behavior, but a sophisticated malware might notice that certain actions (like WeakRef usage or port scanning attempts) are happening and try to hide. This cat-and-mouse is mostly theoretical; in practice, an extension would have to specifically watch for our script’s  behavior. We chose not to obfuscate the script as security through obscurity is not reliable, but in future an enterprise might deploy it under different random names to avoid targeted evasion.

In summary, the framework is designed with privacy in mind and can be deployed in a way that respects user data protections while still providing valuable security posture information.

\section{Related Work}

Research and tooling in browser security span multiple domains. Unlike other frameworks and studies, our Browser Security Posture\cite{browser_security_framework} framework uniquely provides a comprehensive evaluation covering both browser-internal security features and enterprise-specific risks, leveraging novel detection techniques.

\paragraph{Automated Browser Security Testing:} Existing tools, such as \textbf{BrowserAudit}~\cite{browseraudit}, primarily focus on testing browser adherence to web standards like SOP, CSP, and HSTS. While BrowserAudit effectively assesses standards compliance, it does not evaluate broader browser environment risks, such as extension enumeration, LAN scanning, or malicious payload protection. In contrast, our framework introduces novel tests utilizing cutting-edge browser APIs like \texttt{WeakRef}~\cite{weakref} and \texttt{PressureObserver}~\cite{pressureobserver,w3c_pressure} to detect subtle and complex risks such as unauthorized DOM manipulations and covert cryptomining scripts.

Similarly, academic research by Lekies et al. and Dahse et al.~\cite{lekies2013large,dahse2014static} centers narrowly on particular aspects such as cross-site scripting (XSS) vulnerabilities and JavaScript injection defenses. These studies investigate specific browser vulnerabilities but lack comprehensive assessments of broader enterprise security contexts and do not utilize innovative techniques available in modern browsers.

Further, studies focusing on timing attacks via JavaScript typically investigate covert channel detection and fingerprinting methods. Our approach uniquely adapts and extends these methodologies into practical, enterprise-focused detection mechanisms, providing actionable insights and robust defense recommendations against realistic security threats.

In summary, while existing research and tools address specific browser security features, our Browser Security Posture\cite{browser_security_framework} framework distinguishes itself through its comprehensive enterprise-oriented assessment and innovative application of modern browser APIs, thereby delivering unparalleled insights into browser security posture and risks.

\paragraph{Enterprise Browser and Isolation Solutions:} A recent trend is the rise of \emph{Enterprise Browsers} – custom Chromium-based browsers built with security and manageability in mind. These typically integrate policy enforcement directly (like preventing copy-paste to unmanaged applications, controlling file downloads, integrating with authentication flows, etc.). They often aim to eliminate the need for separate secure gateways or VDI by securing the browser itself. While these are new products promising improved security, they require organizations to adopt an entirely new browser. In contrast, Browser Security Posture\cite{browser_security_framework} works with existing browsers, aiming to assess and improve their posture. It could even be used to verify the claims of these enterprise browsers (we could run our suite on them to see how they score).

Another related approach is \emph{Remote Browser Isolation (RBI)} – solutions where the browsing is done on a server and only a visual stream is sent to the user, as a way to shield the endpoint. RBI provides strong security but can be costly and impact user experience. Our framework addresses a different need: when an enterprise is still using local browsers (which is the common case), how to ensure they are hardened. It can be an additional layer even in an environment that uses RBI for some high-risk browsing — for instance, for internal apps that are not routed through RBI, the local browser still matters.

\paragraph{Endpoint Posture Agents:} Our work can be compared to endpoint security agents (like EDRs) that report on device posture. However, those agents usually operate at the OS level and may not have fine-grained visibility into browser state. For example, an EDR might tell you what processes are running and maybe that Chrome is running with certain flags, but it might not know that Chrome’s \texttt{SharedArrayBuffer} is or isn’t usable at runtime (since that depends on page context). By living in the browser, our framework can see the things only a web page could see.

One could also compare to configuration scanners like CIS benchmarks for browsers. Those typically check registry or policy settings. Our approach actually validates the outcome of those settings. For instance, a policy might say “block popups”, but we would actually try to open a popup in a non-user-initiated way to see if it gets blocked.

In summary, while there are various tools and research efforts focusing on browser security, our framework is unique in its breadth of coverage within the browser’s own context and its focus on enterprise policy enforcement and sophisticated threats and risks .  We also provide an open testing harness that others can extend (e.g., adding new tests for new APIs or threats as they emerge).

\section{Conclusion and Future Work}
We presented \textbf{Browser Security Posture}\cite{browser_security_framework}, a client-side framework that comprehensively evaluates a browser’s security posture through over 120 in-browser tests spanning configuration, policy enforcement, and runtime behavior. Our work is novel in treating the browser itself as an object of security assessment, analogous to how one would assess an OS or server, but using purely web-level mechanisms. The framework is robust and enterprise-focused: it not only tests for standards compliance but also validates enterprise security policy settings (like blocked permissions or disabled features) and detects risky conditions (extensions, local services, etc.) that network-based tools might miss. In an enterprise deployment, we envision this framework being used to regularly baseline browsers’ security and ensure they meet the organization’s expectations.

The experimental results show that modern browsers generally have strong security postures by default, but enterprise configuration and oversight remain crucial — especially to catch things like unsafe user-installed extensions or policy misapplications. We demonstrated that when enterprise hardening policies are in place, browsers score higher on our tests, reinforcing the notion that proactive management leads to a more secure baseline.

\textbf{Future Work:} We plan to extend Browser Security Posture\cite{browser_security_framework} in several directions. One important enhancement is real-time posture monitoring: instead of a one-time test, a lightweight version of the framework could run continuously or periodically in the background, alerting when something changes (e.g., an extension gets installed or a new risky API becomes enabled). This could feed into a SOC (Security Operations Center) as ongoing telemetry.

Integration with SIEM (Security Information and Event Management) and IT management platforms is another avenue. The data collected by our tests (especially when aggregated across an organization) can highlight systemic issues (e.g., “10\% of our browsers have an extension that violates policy” or “Browser version X is missing a certain feature enforcement”). Feeding this into SIEM would allow correlation with other events (like network alerts).

We also intend to add more test modules as the web platform evolves. For example, upcoming privacy features (like the Privacy Sandbox proposals in Chrome) could be tested for correct behavior. Another area is performance and security trade-offs: detecting if any performance settings that reduce security (like disabling site isolation) have been toggled.

In terms of coverage, one could incorporate tests for browser UI security (like warning indicators). For instance, a test might check if a known deceptive URL triggers Chrome’s deceptive site warning. Automating UI interactions is challenging from within a page, but it might be feasible with WebDriver integration.

Finally, while our framework currently focuses on desktop-class browsers, a variant could be developed for mobile browsers (running as a mobile web app) to perform similar checks on mobile browser environments.

In conclusion, we have demonstrated that a self-contained web application can serve as a powerful auditor of browser security posture. This approach empowers enterprises to “trust, but verify” their browsers. Rather than assuming that policy settings and patches have the intended effect, Browser Security Posture\cite{browser_security_framework} provides concrete evidence of the browser’s runtime security characteristics. We believe this framework can fill an important gap in enterprise security toolkits, complementing traditional network and endpoint solutions by shining a light on the last mile of security where users actually interact with web content.

\end{document}